\documentclass[conference]{IEEEtran}
\IEEEoverridecommandlockouts
\usepackage{cite}
\usepackage{amsmath,amssymb,amsfonts}
\usepackage{graphicx}
\usepackage{textcomp}
\usepackage{xcolor}
\usepackage{algorithm}
\usepackage{algpseudocode}
\usepackage{listings}
\usepackage{pgfplots}
\pgfplotsset{width=10cm,compat=1.9}
\def\BibTeX{{\rm B\kern-.05em{\sc i\kern-.025em b}\kern-.08em
    T\kern-.1667em\lower.7ex\hbox{E}\kern-.125emX}}
\begin{document}

\title{Circuit Partitioning Using Large Language Models for Quantum Compilation and Simulations}

\author{\IEEEauthorblockN{Pranav Sinha}
\IEEEauthorblockA{\textit{Dept. of Computer Science} \\
\textit{Oakland University}\\
Rochester, MI, USA \\
pranavsinha@oakland.edu}
\and
\IEEEauthorblockN{Sumit Kumar Jha}
\IEEEauthorblockA{\textit{School of Computing and Information Science} \\
\textit{Florida International University}\\
Miami, FL, USA \\
jha@cs.fiu.edu}
\and
\IEEEauthorblockN{Sunny Raj}
\IEEEauthorblockA{\textit{Dept. of Computer Science} \\
\textit{Oakland University}\\
Rochester, MI, USA \\
raj@oakland.edu}
}

\maketitle

\begin{abstract}
We are in the midst of the noisy intermediate-scale quantum (NISQ) era, where quantum computers are limited by noisy gates, some of which are more error-prone than others and can render the final computation incomprehensible. Quantum circuit compilation algorithms attempt to minimize these noisy gates when mapping quantum algorithms onto quantum hardware but face computational challenges that restrict their application to circuits with no more than 5-6 qubits, necessitating the need to partition large circuits before the application of noisy quantum gate minimization algorithms. The existing generation of these algorithms is heuristic in nature and does not account for downstream gate minimization tasks. Large language models (LLMs) have the potential to change this and help improve quantum circuit partitions. This paper investigates the use of LLMs, such as Llama and Mistral, for partitioning quantum circuits by capitalizing on their abilities to understand and generate code, including QASM. Specifically, we teach LLMs to partition circuits using the quick partition approach of the Berkeley Quantum Synthesis Toolkit. Through experimental evaluations, we show that careful fine-tuning of open source LLMs enables us to obtain an accuracy of 53.4\% for the partition task while over-the-shelf LLMs are unable to correctly partition circuits, using standard 1-shot and few-shot training approaches.
\end{abstract}
\vspace{-0.9em}
\begin{IEEEkeywords}
quantum computing, LLM, partitioning
\end{IEEEkeywords}

\section{Introduction}
Quantum circuit compilation converts a quantum algorithm written in a high-level language into elementary quantum gates supported by the quantum hardware. This conversion consists of multiple steps, including first converting the high-level language into intermediate languages (QASM) before mapping onto the quantum computing hardware. Different quantum computing hardware supports different gates, and as can be expected, the mapping from QASM, which is written using generalized gate sets, to the quantum hardware is an involved and complicated process. 

A specific challenge is presented by the fact that right now, we are in the midst of the noisy intermediate-scale quantum (NISQ) era~\cite{Preskill2018quantumcomputingin}. The current-era quantum computers have qubit sizes in the 1000s and are characterized by noisy quantum gates, some gates being more noisy and error-prone than others. For example, the two-qubit CNOT gate is noisier than other gates, and excess CNOT gates in the compiled circuit can render the output incomprehensible. Multiple compilation algorithms focused on minimizing the number of noisy gates have been proposed in the literature~\cite{davis2020towards}.

These compilation algorithms have computational complexity that is exponential in the number of qubits and cannot be directly applied to circuits larger than 5-6 qubits. Therefore, multiple partitioning methods to break down a large circuit into smaller sub-circuits for compilation have been proposed. The problem of partitioning is equivalent to rearranging the intermediate-level QASM quantum code that enables the best application of the downstream compilation algorithm.

Large language models such as GPT and Llama excel at understanding and generating code and have a good knowledge of the QASM code that the partition algorithms work on~\cite{openai_chatgpt,meta_llama3}. There is an expectation that they will perform well at the partitioning task, and investigating this expectation is the main focus of this paper. Specifically, we focus on teaching quick partition, a popular partitioning technique, to large language models~\cite{osti_1785933}. To the best of our knowledge, we are the first to investigate the performance of large language models for partitioning quantum circuits. The paper aims to answer the question of whether a large language model can learn an existing quantum circuit partition algorithm and partition a quantum circuit correctly and, if the circuit is not partitioned correctly, whether the generated code is still correct or not. Through thorough experimental evaluations, we answer the question in the affirmative for both of the questions while making the following observations: 

\begin{itemize}
    \item Even with an expert understanding of the QASM programming language, partitioning is challenging for SOTA LLM models. Off-the-shelf models cannot partition using popular few-shot approaches, and acceptable performance is only achieved after carefully fine-tuning the models. 

    \item The context size of the LLMs is important for processing large circuits, and earlier models with smaller context sizes do not work well as they can only be trained using small quantum algorithms.

    \item Overall, the best accuracy of 53.4\% is obtained by fine-tuning the open-source Llama-3.1 70B model. Accuracy achieved from over-the-shelf LLMs is 0\%.  

    \item Even in cases when the generated code from the fine-tuned LLM is not equivalent to the code generated by quick partition, the partitioned code is equivalent to the original code in 100\% of instances. 
\end{itemize}

% \newpage

\section{Related Work}

\subsection{Quantum Circuit Partitioning}
The NISQ quantum devices are characterized by a number of errors, including state preparation and measurement (SPAM) and qubit state decoherence errors. Two qubit gates like CNOT are more susceptible than one qubit gates and have an error rate that is 1-3\% higher than one qubit gates. This issue of error on quantum computers is well known, and consequently, there is a line of work dedicated to optimizing and reducing the number of gates in quantum circuits. The popular QisKit compiler has four levels of optimization from 0 to 3, with level 3 providing the most optimization~\cite{qiskit2024}. These optimizations include finding logical reductions and commutative possibilities to decrease the gate count. The optimization also uses the noise information of the hardware to select a layout with a favorable noise profile. Patel et al. showed that even after applying all of the optimizations in the QisKit compiler, running TFIM (Transverse Field Ising Model) and Heisenberg quantum algorithms on the relatively low error IBMQ Manila quantum computer, the quantum computer generated results with errors large enough not to provide any meaningful insights~\cite{patel2022quest}. Existing CNOT gate reduction and optimization in QisKit and other compilation software has shown to be insufficient, and multiple works targeting the reduction of CNOT gates have been proposed in the literature.   

Davis et al. proposed the QSearch algorithm to decrease the number of CNOT gates~\cite{davis2020towards}. They build the circuit incrementally and use the A$^*$ search strategy along with the Hilbert–Schmidt norm to find fitness and obtain the best solution. The search has exponential complexity in the number of qubits and can only work with circuits of up to 4 qubits effectively. The QGo algorithm introduced by Wu et al. introduced the partitioning of circuits into blocks of 4 qubits to apply the QSearch algorithm~\cite{wu2020qgo}. They proposed the scan partitioner, a greedy heuristic algorithm to find blocks of a certain qubit size that have the highest number of CNOT gates, with the rationale being that the more CNOT gates in a block, the better the reduction using QSearch. After the optimizations are completed, the blocks are merged together to obtain the equivalent of the original circuit.

Smith et al. proposed the LEAP algorithm, which improves upon QSearch for multiple metrics~\cite{smith2023leap}. LEAP is faster, can decrease CNOT gate count by up to 48$\times$ compared to QSearch, and can scale to 6 qubits. The circuit has to be partitioned similarly to QGo for larger circuits with more qubits. Patel et al. introduced the QUEST algorithm for circuit synthesis that produces approximate instead of exact circuits of the previous approach~\cite{patel2022quest}. This leads to an even greater decrease in the CNOT count, and when the synthesized approximate circuit is run on an error-prone quantum computer, the circuit produces outputs that are more accurate than the previous methods using equivalent circuits. QUEST uses the scan partitioner of earlier methods with a block size of 4 qubits.

\begin{figure}[t]
\includegraphics[]{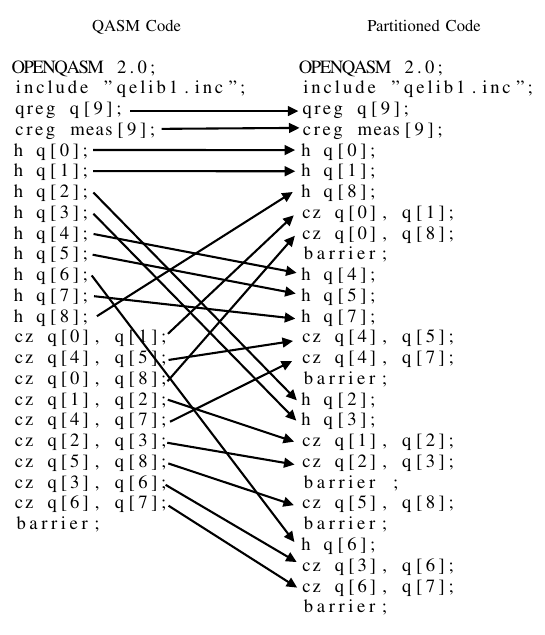}
\caption{QASM code of the benchmark \textit{graphstate indep} of MQT Bench and the corresponding partitioned code. The partitioned code shows significant rearrangement from the original code. The QASM code has been generated from the original benchmark where the gates have been written in the order of their execution. The qubit locations have been removed from the barrier instructions to decrease the LLM token count.}
\label{fig:code}
\end{figure}

Apart from finding efficient mappings of quantum gates to run on the quantum hardware, quantum circuit partitioning is also a critical component for quantum circuit simulation (QCS) on traditional hardware~\cite{arute2019quantum,li2020eliminating,zhang2024quantum}. QCS also has the inherent problem of exponential complexity, and simulating big circuits is slow and infeasible in most cases~\cite{markov2020massively}. Methods such as Pareto-Efficient Quantum Circuit Simulation utilize a partitioning strategy to simulate large circuits~\cite{pednault2017pareto}. Their approach partitions the quantum circuit into subcircuits that can be simulated independently, and then they recombine them to get the desired output. The splitting of the circuit is not arbitrary due to quantum entanglement and has to be done carefully to enable memory and computation efficiency. 

Overall, quantum circuit partitioning is important for both running the code on a quantum computer and simulating it on a classical computer. Different applications require different partitioning schemes, but the partition schemes developed so far have been based on heuristics with no clear justification for the heuristics. There is a need for a more focused approach that partitions the blocks based on the downstream algorithm that is being run on the blocks. LLMs have the potential to identify structures in the circuit that will be better optimized by synthesis algorithms. In this paper, we make a small effort towards that goal and try to see if LLMs can emulate an already existing partition algorithm. Specifically, in this work, we use the quick partitioning scheme, a popular method used in the Berkeley Quantum Synthesis Toolkit (BQSKit), and try to simulate it using LLMs~\cite{osti_1785933}. An algorithmic description of quick partition is provided in the methods section and an example of quick partitioning is shown in Figure~\ref{fig:code}.

\subsection{Large Language Models for Coding}
A multitude of recent advances in neural network architecture and training have come together to enable the incredible performance of large language models (LLMs). The transformer neural network architecture~\cite{vaswani2017attention} has superseded the recurrent neural network (RNN)~\cite{rumelhart1986learning} and long-short-term memory (LSTM) architecture~\cite{hochreiter1997long} for natural language processing by providing non-sequential processing for previously sequential operations. The generative pre-training (GPT)~\cite{radford2018improving} training regime has enabled training large amounts of unlabeled text data available online to create LLMs that are experts at a variety of language-related tasks.

The most well-known LLMs are from the GPT family of models from OpenAI~\cite{openai_chatgpt}. These models range from GPT-1 to GPT-4, and all have transformer architecture. Subsequent models in the series have been trained on larger datasets and have more trainable parameters. Though details of the architecture have not been made public, the latest GPT-4 model is thought to have more than a trillion parameters. The large size and the closed-source nature of GPT-4 make them difficult to run locally and have prompted the development of multiple open-source alternatives. 

The Llama family of models has been developed by the Meta group to optimize the inference budget instead of the training budget optimized by most other LLMs~\cite{touvron2023llama}. It is achieved by training the models for longer than typical and can lead to smaller, better-trained LLMs, which are cheaper to run during inference. The Llama 1 model has been trained on 1.4T tokens consisting of CommonCrawl (67\%)~\cite{commoncrawl}, C4 (15\%)~\cite{raffel2020exploring}, Github (4.5\%), Wikipedia (4.5\%), Gutenberg and Books3 (4.5\%)~\cite{gao2020pile}, ArXiv (2.5\%), and Stack Exchange (2\%). Llama 1 uses the transformer architecture with a few changes, such as using pre-normalization, SwiGLU~\cite{shazeer2020glu} activations instead of ReLU, and rotary embedding~\cite{su2024roformer}. The Llama 3.1 models have been trained using better-quality data and are larger in size. The Llama 3.1 models come in various sizes, 8B, 70B, and 405B, and the larger 405B model has been trained on 15.6T text tokens~\cite{dubey2024llama}. These models provide impressive zero and few-shot performance on  Code, Math, Reasoning, Tool Use, and Multi-Lingual and Massive Multitask Language Understanding (MMLU) benchmarks~\cite{hendrycks2020measuring}. 

Mistral AI has also released a diverse model family that excels at code and reasoning, tool use, and instruction following~\cite{mistral2024}. Mistral Large 2 has 123B parameters and has been designed for large context sizes. Mixtral 8x7b and Mixtral 8x22b are sparse Mixture-of-Experts (SMoE) models and have been designed for inference efficiency. Mixtral 8x22b consist of 8 22B models, but only use a subset during inference for energy efficiency. All of the above mentioned models are also based on the transformer architecture. 

Multiple coding-focused LLMs have been developed to provide assistance while coding. The above-mentioned SOTA LLMs have all been trained on coding data from GitHub, Stack Exchange, and other sources but have a focus on improving general language performance. The coding LLMs, on the other hand, have a primary focus on writing and understanding common programming languages. CodeLlama, for instance, is built on top of Llama 2, but more of the training time has been devoted to code, and as a result, it performs better at coding compared to its base model~\cite{roziere2023code}. Code Llama-Python is specifically trained in writing and understanding Python code. Another critical feature of coding LLMs is context size. Some codes can be large and require more number of tokens to make sense of the complete functionality. The Codestral Mamba model from Mistral AI uses the Mamba architecture instead of transformers, can provide inference in real-time, and can theoretically model sequences of infinite lengths~\cite{gu2023mamba}. 

\begin{figure}[t]
    \centering
    \includegraphics[width=1.0\linewidth]{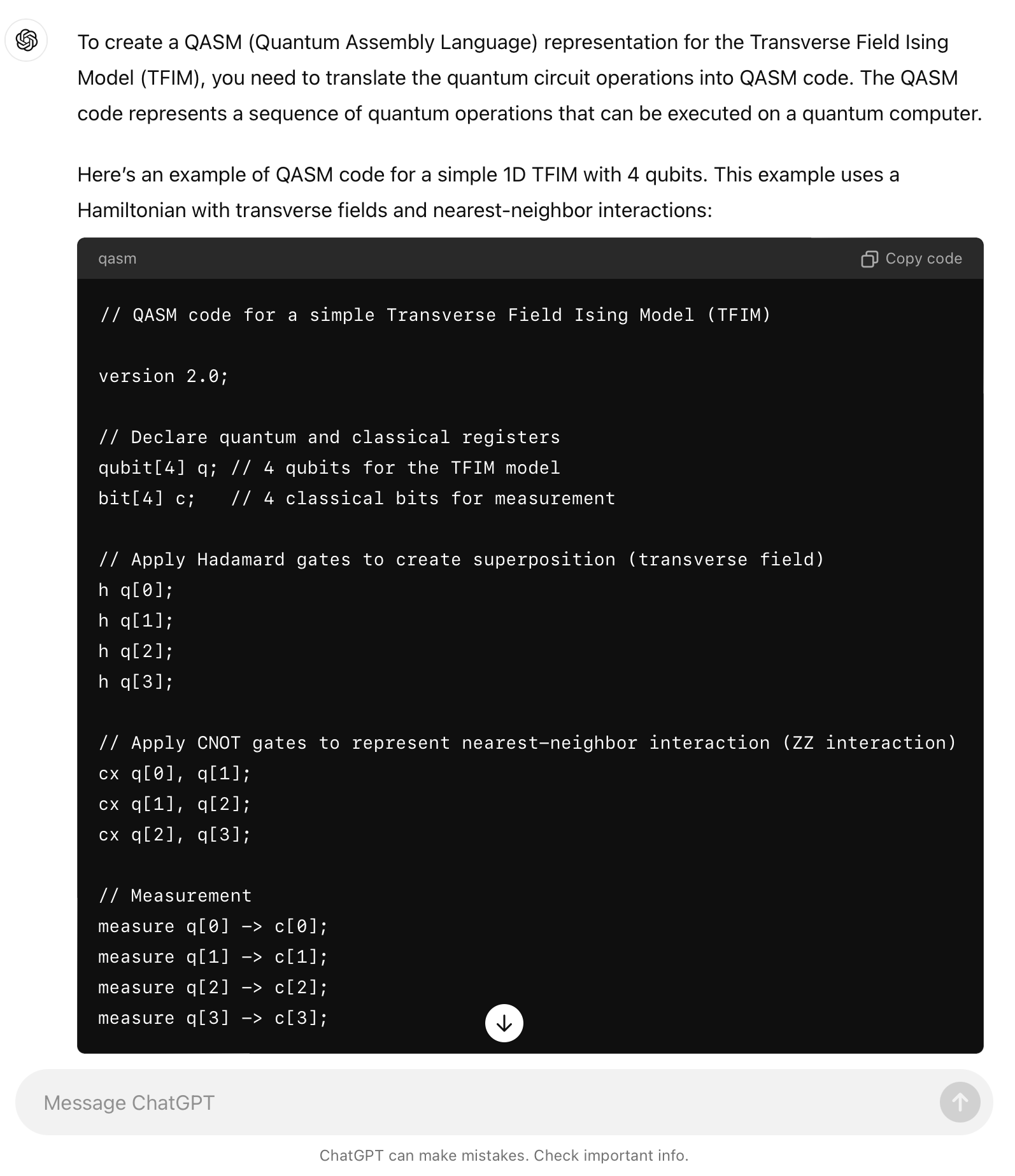}
    \caption{ChatGPT-4o mini producing QASM code for TFIM. It can understand and write QASM code but is unable to partition the code into blocks.}
    \label{fig:chatgpt}
\end{figure}

A report containing metrics on the use of quantum programming languages and QASM has yet to be published for all of the above-mentioned LLMs. Nevertheless, in our experience, most of the models can understand and write quantum code. Figure~\ref{fig:chatgpt} shows the output generated by ChatGTP-4o mini when it is asked to write a 4-qubit TFIM QASM code, demonstrating its understanding of QASM and quantum circuits. This proficiency could have been acquired from Wikipedia, Github, Stack Exchange, and ArXiv text used to train the models. In this paper, we are focused on partitioning a quantum circuit written in QASM, and this is something that the SOTA LLMs have not been exposed to before. LLMs are quick learners, and 1-shot or few-shot learning is sufficient to learn various tasks and is the primary way of using LLMs. In this paper, we show that this approach is insufficient for getting good performance for partitioning tasks, and as a result, we have to fine-tune the whole model to get good performance.

\section{Method}

\subsection{Data Preparation}
To fine-tune the LLM for the partitioning task, we need quantum circuits and their equivalent partitioned circuits. The quantum circuits were obtained from the Munich Quantum Toolkit Benchmark Library (MQT Bench); we used all of the fixed benchmarks and the circuits of sizes 2-100 of the scalable benchmark~\cite{quetschlich2023mqt}. The LLMs have a limit on the number of tokens that they can process, specifically given the limited graphics memory on consumer GPUs. We decrease the number of tokens from the code by removing comments and measure instructions. We then run quick partition on the cleaned code. The partitioner outputs multiple separate blocks; we combine the blocks together by adding a barrier between the blocks to create a single file. Finally, the QASM code consists of multiple floating point numbers; during our initial experiments, we observed that these floating points were utilizing extra tokens, and the LLM had difficulty reproducing the floating point values. We replaced the tokens with symbols in both the input and the output file. This transformation has a further advantage of decreasing the size of the quantum code.

\begin{figure}[t]
    \centering
    \begin{tikzpicture}[scale=0.8]
\begin{axis}[
            symbolic x coords={3000, 6000, 12000, 24000, 48000, 48000+ },
            width=8cm,
            bar width=0.6cm,
            ylabel={No. of circuits},
            xlabel={No. of tokens},
            %nodes near coords,
            %nodes near coords align={vertical},
            xtick=data
          ]
            \addplot[ybar,fill=gray] coordinates {
                (3000,349)
                (6000,136)
                (12000,98)
                (24000,119)
                (48000,142)
                (48000+,475)
            };
        \end{axis}
\end{tikzpicture}
    \caption{Distribution of the number of circuits in relation to the number of tokens in the MQT Bench dataset. The number of tokens are for Llama3.1.}
    \label{fig:histogram}
\end{figure}
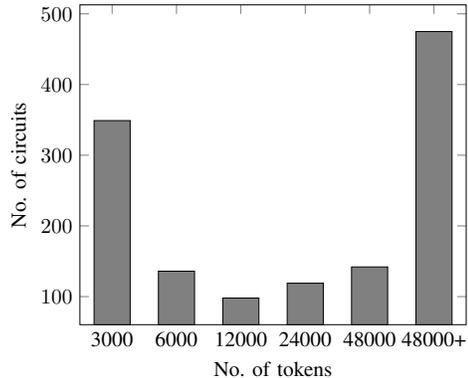

Once all of the pre-processing transformations were applied, we still had to deal with the constraints on the number of tokens that could be used during fine-tuning. Every LLM has an associated way of obtaining tokens from the text input, and there is a limit on the context size of the LLM, which is the number of tokens that the LLM can look at one time. There is a further limit on the number of tokens that can be used to train locally, as more tokens require more GPU memory. For example, Llama 3 has a context size of 8000. However, we could not utilize all of the context sizes on our local setup without running out of memory, so we settled on 6000 as the maximum token size of the input. A plot showing the distribution of number of tokens is presented in Figure~\ref{fig:histogram}, and we can observe that a significant number of circuits have a token size of more than 48000, and it is not something that can be used to train the existing SOTA LLMs. A method to utilize all of the data to train the LLM could potentially lead to better partition accuracy.

\subsection{LLM Prompts}
LLM prompts are critical to the success of the results obtained from LLM. The prompt that we used is as follows: \textit{``Create barriers for efficient processing. Make sure that you check gates with multiple qubits and do not change their order if they depend on each other:"}. The first sentence provides the main goal of the fine-tuning, which is partitioning the circuit into blocks separated by barriers in the training data. The second sentence was needed as we observed the LLM mixing up the order of the gates within the block. While fine-tuning the LLM, we also added \textit{``End of barrier creation"} at the end of each input. LLMs can be chatty and sometimes continue to generate extra code; we added the line to counter that. If the LLM generated any extra code after \textit{``End of barrier creation,"} we ignore that code and do not use it to compare it to the correct code. 

\subsection{Quick Partition}
The BQSKit quantum synthesis toolkit implements several partitioners, including clustering, greedy, scan, and quick partitioner~\cite{osti_1785933}. The quick partitioner is currently recommended as the default partitioner. It is a simple partitioner that partitions the circuit in one pass and does not specifically optimize any higher-level objective. Although it is the default and the best partitioner in most cases, it has yet to be described in the literature. We remedy that and provide an Algorithm for quick partition (Algorithm 1).

The algorithm takes a circuit $C$ as input. It then initializes an empty partitioned circuit $P$ and a list of active partitions $A$. Active partitions are partitions that have not been added to $P$ and still have room to add gates to them. The loop (line 3) iterates through the gates $G$ in the order of their execution on the quantum hardware. Some gates can be executed in parallel, in which case the lowest qubit index of the gate determines the order. A list of active partitions $SQ$ that share qubits with the gate $G$ is then created. This list is further refined into partitions $ AQ $ that can accommodate $ G $. This list is determined by ensuring that only those partitions from $SQ$ are selected that do not increase the number of qubits from a pre-determined value. If $AQ$ is empty, then a new partition $NP$ is created and added to the list of active partitions. If it is not, then the most suitable pre-existing partition is assigned to $NP$. The steps to determine the most suitable partition are as follows: if the qubits of the gate $G$ is a subset of the qubits of a partition in $AQ$, then it is considered the best partition and selected; otherwise, the first partition in $AQ$ is selected as the suitable partition. The gate $G$ is then added to the $NP$ partition. The qubit is then tagged as blocked in the rest of the active partitions so that no new additions to that qubit are made in another active partition. This blocking can lead to some partitions in $A$ that have all qubits blocked and, consequently, are unable to add any new gates. These now inactive partitions are collected in the $PB$ list and merged with the partitioned circuit. The for loop (line 17) iterates through the partitions $CB$ in $PB$. The qubit locations of the $CB$ are extracted and then checked to find if there are partitions $EB$ at the end of the partitioned circuit $P$ that are subsets of the qubit locations of $CB$ or the other way around. If it is, the $CB$ and $EB$ blocks are merged and added to the partitioned circuit. If no such partition is found in $P$, then $CB$ is added to $P$ without merging it into an existing partition. Finally, after all of the gates are processed (line 26), any remaining active partitions are merged or added using the same operations as in lines 18-23 to the partitioned circuit $P$. 

\begin{algorithm}[t]
\caption{Quick Partition}\label{alg:qp}
\begin{algorithmic}[1]
\Require Circuit $C$
\Ensure Partitioned circuit $P$
\State Initialize partitioned circuit $P$
\State Initialize active partitions $A$
\For {gate $G$ in $C$}
\State $SQ \gets$ active partitions sharing qubits with $G$
\State $AQ \gets$ partitions from $SQ$ that can add $G$
\If{$AQ$ is empty}
\State Create a new partition $NP$ 
\State Add $NP$ to $A$
\Else  
\State $NP \gets$ most suitable partition from $AQ$
\EndIf
%\State $SQ \gets$ most suitable block from $AQ$ or new block if $AQ$ is empty
%\State Add $SQ$ to $A$ if $SQ$ is new
\State Add the gate $G$ to the partition $NP$
\State $q \gets$ locations of qubits in $SQ$
\State $EB \gets$ partitions at the end of $P$ 
\State Block additions to qubit $q$ from all active partitions
\State $PB \gets$ partitions blocked on all qubits 
%\State \Comment{Merge closed bins to blocks in partitioned circuit $P$}
\For {$CB$ in $PB$}
\State $q \gets$ qubits in $CB$
\If{$q$ is a subset of a partition $EB$ in $P$ or vice-versa}
\State Merge $CB$ and $EB$
\Else
\State Add $CB$ to $P$
\EndIf
\EndFor
\EndFor
\State Close all open partitions and merge or add to $P$
\end{algorithmic}
\end{algorithm}

\subsection{LoRA}
Our goal in the paper is to investigate the partitioning capabilities of multiple LLMs and find the best accuracy and the best LLM architectures available for them. Since 0-shot, 1-shot, and Few-shots do not work for partitioning, we have to retrain or adapt the LLMs for the partitioning task. Our dataset of quantum programs is small compared to the size of the LLMs, so retraining is not feasible, leaving adapting the trained network for the job as the suitable option. There are two commonly used strategies for adaptions: one is adding an adaption layer at the end of the network, and the other is fine-tuning the whole network. For the first approach, during the retraining process, only the last layer of the network has to be retrained, and it is faster than fine-tuning, where all of the weights of the network have to be updated. Even though fine-tuning is slower and more cumbersome, it gets better results and is our method of choice. Fine-tuning a complete LLM requires a lot of GPU memory, and it is infeasible to do it on consumer-grade GPUs. Instead, we opt for low-rank adaption of large language models (LoRA) to fine-tune the transformers~\cite{hu2021lora}.

The LoRA method freezes the original weights of the LLM and adds an adapter layer composed of the rank-decomposition matrix of the layers. Consequently, a single large layer of the LLM is broken down into two smaller matrices that have a combined lower memory requirement. The training is only carried out on the smaller matrices, and as a result, the LoRA method can reduce trainable parameters by 10,000 times and GPU memory requirements by three times. For a pre-trained weight layer $W_0 \in \mathbf{R}^{(x\times y)}$, during training for iteration $i$ the weight has to be updated by adding $\Delta W$ to it to get the new set of weights $W_0^{i+1} = W_0^i + \Delta W$. In the LoRA method the $\Delta W$ is replaced by the rank decomposition $BA$, where $B\in\mathbf{R}^{x\times r}$ and $A\in\mathbf{R}^{r\times y}$, and the rank ($r$) $<<$ min$(x,y)$. During training, only the small A and B matrices are updated instead of the large $\Delta W$ matrix. The forward pass using the decomposed matrices is calculated as follows: $o = W_0i + BAi$, where $o$ is the output of the layer, and $i$ is the input to the layer, thereby reducing the total retraining memory requirements.

In the transformer architecture, there are four sets of weights in the multi-head attention named $W^V$, $W^K$, $W^Q$, and $W^O$. There are an additional two sets of feed-forward network weights, one each in the encoder and the decoder layer. We fine-tune all six sets of weights for the partitioning task. The rank $r$ is another tunable hyper-parameter that determines the number of training parameters. A higher value of $r$ increases the number of trainable parameters and consequently increases the GPU memory requirements for fine-tuning. Larger $r$ generally gives better performance, but there is a point of diminishing return after which there are no significant performance gains. Using LoRA on GPT-3 with $r=4$ and fine-tuning only the multi-head attention weights, it has been shown that the training checkpoint size decreased from the original 350GB to 35MB. All of the LLMs we used for our experiments are based on the transformer architecture but have small modifications and can have some extra sets of trainable weights. If available, we include any extra weights for fine-tuning the LLM.

\section{Experiments}
We run our experiments on a system consisting of 8 Nvidia H100 GPU with 80 GB of graphics memory each. The first set of experiments test the performance of open source over-the-shelf LLMs and compare it to our LoRA fine-tuning approach. We first found all of the circuits with a token size of less than 6000, and then trained the circuit on 80\% of the dataset and kept the remaining for 20\% for testing. The Llama-3.1 8B model was trained using the standard 1-shot and 5-shot approaches with the following prompt for 1-shot training: 

\textit{``Create barriers for efficient processing. Make sure that you check gates with multiple qbits and not change their order if they depend on each other. Here is one example: \textbackslash n $<$U1$>$ \textbackslash n - - - \textbackslash n Answer: $<$P1$>$ \textbackslash n End of barrier creation. Create partitions for this: \textbackslash n $<$U2$>$ \textbackslash n - - - \textbackslash n Answer: "} %\textit{``Create barriers for efficent processing. Make sure that you check gates with multiple qbits and not change their order if they depend on each other. Here is one example: \textbackslash n {E1[`input']}\textbackslash n - - - \textbackslash n Answer: {E1[`output']} \textbackslash n End of barrier creation. Create partitions for this: \textbackslash n {E2[`input']} \textbackslash n - - - \textbackslash n Answer: {E2[`output']} \textbackslash n End of barrier creation."}

In the above text, U1 is replaced with the original unpartitioned QASM code, while P1 is the corresponding partitioned code with barriers in it. The  U2 is the test code, and the LLM is supposed to generate the partitioned code corresponding to it. The 5-shot training had five examples before \textit{``End of barrier creation"}, instead of one. Few-shot training is the most popular strategy for training LLMs as it is straightforward to implement and requires significantly fewer resources compared to LoRA-based fine-tuning. The unpartitioned text has all of the transformations mentioned in the methods section to reduce the token size. For comparison, we fine-tune the same model using the parameter values mentioned in the methods section for 10 epochs and a learning rate of $1\times10^{-4}$. The results of the experiments are shown in Table~\ref{tab:few_shot}.

\begin{table}[thbp]
    \caption{Few-shot partition accuracy using Llama-3.1-8B. The LLM cannot partition using 1-Shot and 5-Shot methods and repeats the input code. In contrast, using the fine-tuning approach does not lead to repetitions of the input.}
    \label{tab:few_shot}
    \centering
    \begin{tabular}{c|c|c|c}
    \hline
    Method & Repeated Code & Correct Code & Accuracy\\
    \hline
        1-Shot &  95\% & 95\%& 0\%\\
        5-Shot &  100\% & 100\% & 0\%\\
        Fine-tuning  & 0\% & 100\% & 51.69\%\\
        \hline
    \end{tabular}

\end{table}

We observe that few-shot training is inadequate for training LLMs to partition the QASM code. The 1-shot and 5-shot strategy repeats 95\% and 100\% of the input respectively without applying any barriers to the code. LLM fine-tuned using our proposed approach does not repeat the input and is able to produce the same output as a quick partitioner 51.7\% of the time. Our proposed approach partitioned code that is correct 100\% of the time. The equivalence is checked by removing the partitions and recalculating the operation cycle. The 1-shot and 5-shot approaches produce the correct code 95\% and 100\% of the time respectively, but this comparison with our proposed approach is without merit as the LLM simply repeats the correct input.

We also train multiple different types of LLMs, including Phi-3-mini-128k-instruct~\cite{abdin2024phi}, Llama-3.1-8B, Llama-3.1-70B, Mistral-7B and CodeLlama-7B to compare and contrast their performance on the partition task. All of the models were quantized to 8 bits. Different models can have different token size for the same input, we choose 6000 as the maximum token size, and used 80\% for training and 20\% for testing. All of the models use the same training parameters of LoRA, with a rank of 8, a scaling factor of 32, and a dropout of 0.05. All of the trainable layers were trained, which for Llama-3.1 are: q\_proj, up\_proj, o\_proj, k\_proj, down\_proj, gate\_proj, and v\_proj. We train the networks for 10 epochs with a learning rate of $1\times10^{-4}$. Some LLMs can be chatty, so we discard any text produced after the \textit{``End of barrier creation."} line. The run time for each epoch and the accuracy are presented in Table~\ref{tab:llms}. Out of all of the models being run, the lowest accuracy of 15.45\% was obtained using the Phi-3-mini-128k-instruct model. It is the smallest model we tested, containing 3.8 billion parameters; it is also the fastest model to train, taking 25 minutes and 51 seconds per epoch. The best accuracy was obtained from Llama-3.1-70B, the largest model we tested with a total of 70B parameters. Each epoch took around 5 hours to run, and the total training time for all 10 epochs was more than 2 days. Three of the models that we tested have 7-8B parameters and have been designed to run on consumer-grade GPUs; out of these, the Llama-3.1-8B obtained the best accuracy of 51.69\% but also took more time to train compared to the other small models.

\begin{table}[tbp]
\caption{Per-epoch run time and accuracy of various LLMs fine-tuned using our proposed approach.}
 \label{tab:llms}
\begin{center}
\begin{tabular}{c|c|c}
\hline
 Model & Fine-tuning Time Per Epoch & Accuracy \\ 
 \hline
Phi-3-mini-128k-instruct & 0 hours 25 min 51 sec  & 15.45\% \\
Llama-3.1-8B & 0 hours 45 min 36 sec &  51.69\% \\
Llama-3.1-70B & 4 hours 54 min 3 sec &  53.39\% \\
Mistral-7B & 0 hours 38 min 24 sec & 49.09\%\\
CodeLlama-7B & 0 hours 37 min 40 sec &  47.27\% \\

\hline
\end{tabular}
\label{tab1}
\end{center}
\end{table}

\section{Conclusion and Future Work}
We present an LLM-based method for partitioning quantum circuits similar to the quick partition algorithm from the BQS-kit library. Our fine-tuned LLM creates the exact circuit as a quick partitioner 53.39\% of the time. Even when the exact quick-partition circuit is not reproduced, the partitioned circuit is equivalent to the original circuit 100\% of the time. The results presented here are promising and can enable future development of LLMs that are fine-tuned, keeping the various downstream tasks in mind, including directly decreasing the number of CNOT gates. Humans have difficulty designing quantum algorithms and quantum circuits due to their un-intuitive nature. The current research demonstrated that large language models can understand and manipulate quantum circuits to perform non-trivial tasks and have the potential to transform quantum computing.  

\section{Acknowledgments}
We acknowledge support from NSF Award \#2245756 and DOE Award \#DE-SC0024598 to Sunny Raj.
\newpage

\bibliographystyle{IEEEtran}
\bibliography{IEEEexample}

\end{document}